\documentclass[sigconf, nonacm]{acmart}

\usepackage[utf8]{inputenc}
\usepackage{xcolor}
\usepackage{multirow}
\usepackage{tabularx}

\title{PLMGH: What Matters in PLM-GNN Hybrids for Code Classification and Vulnerability Detection}

\author{Mohamed Taoufik Kaouthar El Idrissi}
\email{taoufik.kaouthar-el-idrissi@etud.polymtl.ca}
\affiliation{
  \institution{Polytechnique Montréal}
  \city{Montréal}
  \country{Canada}
}

\author{Edward Zulkoski}
\email{ed@quantstamp.com}
\affiliation{
  \institution{Quantstamp}
  \city{San Francisco}
  \country{United States}
}

\author{Mohammad Hamdaqa}
\email{mhamdaqa@polymtl.ca}
\affiliation{
  \institution{Polytechnique Montréal}
  \city{Montréal}
  \country{Canada}
}

\begin{document}
 
\begin{abstract}
Code understanding models increasingly rely on pretrained language models (PLMs) and graph neural networks (GNNs), which capture complementary semantic and structural information.
We conduct a controlled empirical study of PLM$\rightarrow$GNN hybrids for code classification and vulnerability detection tasks by systematically pairing three code-specialized PLMs with three foundational GNN architectures. We compare these hybrids against PLM-only and GNN-only baselines on Java250 and Devign, including an identifier-obfuscation setting.
Across both tasks, hybrids consistently outperform GNN-only baselines and often improve ranking quality over frozen PLMs. On Devign, performance and robustness are more sensitive to the PLM feature source than to the GNN backbone. We also find that larger PLMs are not necessarily better feature extractors in this pipeline, and that the PLM choice has more impact than the GNN choice. Finally, we distill these findings into practical guidelines for PLM$\rightarrow$GNN design choices in code classification and vulnerability detection.
\end{abstract}

\maketitle

\section{Introduction}
Deep learning is now a core tool in software engineering, supporting tasks such as clone detection, vulnerability detection, code classification, naming, repair, and synthesis~\cite{chen2021evaluating}. Most modern approaches to code understanding fall into two broad families: (i) Pretrained Language Models (PLMs) that process code as a token sequence, and (ii) graph neural networks (GNNs) that operate on structural representations such as abstract syntax trees (ASTs) and program graphs.

PLMs achieve strong accuracy on many benchmarks, largely by learning rich semantic representations from large-scale corpora. However, their practical cost can be substantial: finetuning and inference may require large memory footprints, careful batching, and non-trivial latency, especially when long contexts are needed.

In contrast, GNNs explicitly exploit structure and have shown promise on ASTs and other graph-based representations of code~\cite{allamanis2017learning, leclair2020improved, nguyen2022regvd, zhang2024cross, wei2020lambdanet}. Their structured inductive bias is attractive for program analysis, and their computational footprint is often lower than that of large PLMs, especially when PLMs must be finetuned or run over long contexts. Yet, on standard benchmarks, GNN-only models frequently lag behind strong pretrained PLM baselines, suggesting that structure alone may often not recover the same level of semantic knowledge.

A natural direction is therefore to combine the semantic strength of PLMs with the structural inductive bias of GNNs. Broadly, the literature follows two routes. The first route makes PLMs more structure-aware by injecting program structure into the self-attention mechanism or the pretraining objective, for example through AST-graph-aware attention ~\cite{feng2020codebert, guo2020graphcodebert}, structural positional encodings ~\cite{zhang2023implant}, or explicit structural relations ~\cite{guo2020graphcodebert}. The second route keeps the PLM as a semantic feature source and injects contextual token representations into a downstream graph model ~\cite{yang2024security}, which then performs structure-aware reasoning over an AST or program graph. In this paper, we focus on the second route (PLM$\rightarrow$GNN feature injection) because it is a pragmatic way to reuse strong pretrained representations while keeping the downstream model lightweight and explicitly structure-aware ~\cite{yang2024security}.



Despite growing popularity of PLM$\rightarrow$GNN pipelines, existing studies often evaluate a single PLM with a single graph backbone~\cite{yang2024security}, which makes it difficult to answer four practical questions: (i) do PLM$\rightarrow$GNN hybrids consistently outperform PLM-only and GNN-only baselines, (ii) what computational costs do these hybrids introduce, (iii) how robust are they under identifier obfuscation, and (iv) does performance depend primarily on the chosen PLM feature source or on the GNN architecture. These questions are timely because PLMs are increasingly used as drop-in components in SE pipelines, yet their deployment cost motivates hybrid designs that may preserve accuracy while reducing compute and improving robustness.



\begin{figure*}[!t]
  \hspace*{-1cm}\includegraphics[width=1.15\textwidth]{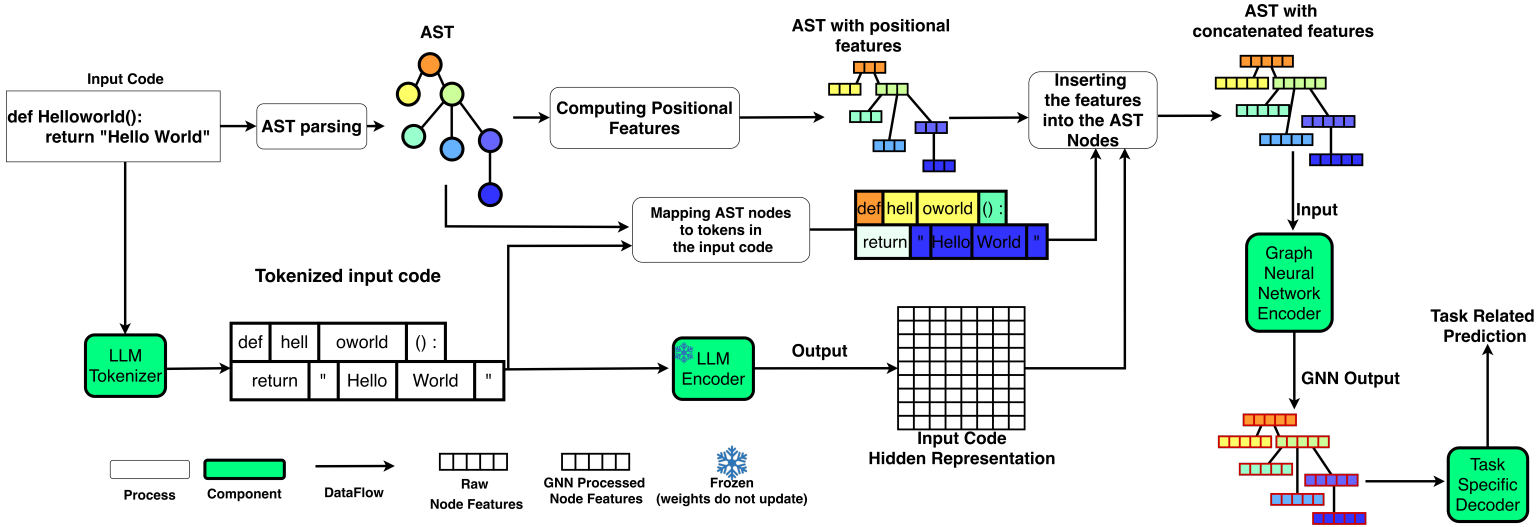}
  \centering
  \caption{Overview of the PLM$\rightarrow$GNN approach}
  \label{fig:inject_diagram}
\end{figure*}

To address this, we conduct a systematic empirical study of PLM$\rightarrow$GNN hybrids for code understanding tasks under a unified and controlled evaluation protocol. We consider three foundational GNN architectures: GCN ~\cite{kipf2016semi}, GAT ~\cite{velivckovic2017graph}, and a GraphTransformer ~\cite{shi2020masked}, spanning convolutional aggregation (GCN), attention-based message passing (GAT), and transformer-style graph reasoning (GraphTransformer). We pair this GNN selection with three recent code-specialized PLMs: DeepSeek-Coder-1.3B ~\cite{guo2024deepseek}, StarCoder2-3B ~\cite{lozhkov2024starcoder}, and Qwen2.5-Coder-0.5B ~\cite{hui2024qwen2}, covering a representative range of model sizes. For each PLM$\rightarrow$GNN combination, we freeze the PLM and align token-level embeddings to AST nodes before processing them with a GNN, isolating the effect of semantic feature injection while keeping training practical across many pairings. We evaluate these hybrids alongside GNN-only and PLM-only baselines (frozen and, where applicable, finetuned) on two widely used benchmarks: Java250 code classification \cite{puri2021codenet} and Devign vulnerability detection \cite{zhou2019devign}, including an out-of-distribution Devign setting with identifier obfuscation. These benchmarks capture complementary settings (multiclass classification and imbalanced binary detection), enabling us to study effectiveness, efficiency, and robustness while keeping the experimental matrix tractable.

Overall, we observe that the choice of PLM feature source tends to have a larger impact on performance than the choice of GNN backbone in our evaluated settings. We additionally find that the PLM size is not the primary factor that decides on the PLM$\rightarrow$GNN hybrid performance. 

This work investigates the following research questions:
\begin{itemize}
  \item \textbf{RQ1 (Effectiveness):} How do PLM$\rightarrow$GNN hybrid models compare to GNN-only and PLM-only baselines in predictive performance across code classification and vulnerability detection tasks?
  
  \item \textbf{RQ2 (Efficiency):} What computational costs do PLM$\rightarrow$GNN hybrids introduce in terms of preprocessing time and inference latency?
  
  \item \textbf{RQ3 (Robustness):} How does identifier obfuscation affect PLM$\rightarrow$GNN performance compared to the in-distribution setting, and how does this compare to PLM-only and GNN-only baselines?
  
  \item \textbf{RQ4 (Design sensitivity):} How do the choices of PLM feature source and GNN family interact, and which design factors most strongly influence performance?
\end{itemize}

Answering these research questions allows us to characterize what drives PLM$\rightarrow$GNN performance and provide practical guidance on designing PLM$\rightarrow$GNN pipelines.
Our contributions are:
\begin{itemize}
  \item A characterization of performance, robustness (including identifier-obfuscation OOD), and compute costs.
  \item Practical guidance on selecting PLM features and GNN families under different constraints.
  \item A controlled empirical comparison of PLM-only, GNN-only, and PLM$\rightarrow$GNN hybrid models across two widely used benchmarks, with consistent data splits and evaluation protocol.
\end{itemize}

The remainder of the paper is organized as follows. Section~\ref{sec:study_design} describes the study design. Section~\ref{sec:results} presents the results. Section~\ref{sec:discussion} discusses key findings, practical guidelines, and threats to validity. Section~\ref{sec:related} reviews related work. Our implementation is available online.\footnote{\url{https://github.com/PlayeerOne/PLMGH}}

\section{Study design}
\label{sec:study_design}

\subsection{Overview}
\label{sec:overview}

Our objective is to assess whether combining pretrained semantic representations from code PLMs with explicit structural reasoning in GNNs improves performance on code understanding tasks compared to using either component alone. We evaluate this trade-off in terms of predictive performance, robustness under identifier obfuscation, and computational cost under a controlled protocol. To isolate the effect of representation transfer and enable fair comparison across many PLM$\times$GNN pairings, we use PLMs as \emph{frozen} feature extractors and train only the structural components (feature fusion, GNN backbone, and classifier). We keep the program representation (AST graphs), data splits, and training budget fixed across methods, while varying (i) the PLM feature source and (ii) the GNN architecture.

Figure~\ref{fig:inject_diagram} summarizes our PLM$\rightarrow$GNN injection pipeline. Given a source file, we (i) parse the program into an AST graph, (ii) extract contextual token representations from a frozen pretrained code PLM, and (iii) align token embeddings to AST nodes to obtain semantic node features. We then fuse injected semantic features with structural node features and apply a GNN followed by a graph-level classifier.

In this paper, \emph{injection} refers to mapping frozen PLM token embeddings (from the last hidden layer) to their corresponding AST nodes via token--node alignment. For each node, we augment injected semantic features with structural features, including positional encodings and node-type embeddings. The PLM parameters remain fixed; only the fusion module, the GNN backbone, and the task classifier are trained end-to-end. Hyperparameters are tuned on Java250 and transferred to Devign to keep the tuning budget comparable across the PLM$\times$GNN grid.


\subsection{Datasets and Tasks}
\label{sec:datasets_and_tasks}
For evaluation, we use two widely used benchmarks that cover complementary code-understanding tasks: \emph{multiclass program classification} and \emph{vulnerability detection}. Java250~\cite{puri2021codenet} provides a controlled large-label-space setting where structural cues matter for problem intent, while Devign~\cite{zhou2019devign} targets security-relevant reasoning at the function level. These benchmarks cover two core code understanding tasks: code classification and vulnerability detection. Together, they allow us to test effectiveness (RQ1), quantify performance--efficiency trade-offs (RQ2), and probe robustness under distribution shift (RQ3), while keeping the experimental matrix tractable for a systematic sweep (RQ4).

Java250 is a multi-class classification dataset derived from Project CodeNet~\cite{puri2021codenet}, where each sample is a Java program labeled by the programming problem it solves. The dataset contains 250 classes with 300 solutions per class, totaling 75{,}000 samples. We use a standard (3:1:1) train/validation/test split and keep it fixed across all models for fair comparison.\footnote{\url{https://github.com/IBM/Project_CodeNet}}

Devign~\cite{zhou2019devign} is a binary classification benchmark for detecting whether a C function is vulnerable. We use the CodeXGLUE release provided by~\cite{TRAN2025107504}, which contains 27{,}318 labeled functions extracted from security-related commits in two large real-world projects (QEMU and FFmpeg). The dataset is randomly shuffled and partitioned into train/validation/test using an 8:1:1 ratio.

To evaluate robustness (RQ3), we use the out-of-distribution (OOD) test set from~\cite{TRAN2025107504} which applies identifier obfuscation and is available on HuggingFace.\footnote{\url{https://huggingface.co/datasets/DetectVul/devign}} Concretely, for each function, user-defined identifiers (variable names, function names, type identifiers) are replaced with canonical placeholders in a consistent manner within the function, while preserving keywords, operators, literals, and formatting required for parsing. Models are trained on the original (non-obfuscated) Devign training set and evaluated on both the standard Devign test set and the obfuscated Devign-OOD test set.

The ground-truth targets are the benchmark labels provided with each dataset. In Java250, each program is assigned one of 250 class labels corresponding to the programming problem it solves. In Devign and Devign-OOD, each function is assigned a binary label indicating whether it is vulnerable or non-vulnerable. All reported classification metrics compare these predicted labels against the benchmark ground-truth labels on the held-out test sets.

\subsection{Models and Baselines}
\label{sec:models_and_baselines}
We evaluate three model families under a unified pipeline and training protocol.

\textbf{GNN-only baselines :}
For each AST graph, nodes receive structural features only: a node-type embedding and a positional encoding (Section~\ref{sec:program_representation}). These features are fused into node vectors (Section~\ref{sec:program_representation}) and processed by a GNN followed by a graph-level classifier. Unless stated otherwise, we fix core GNN capacity to 8 message-passing layers and for attention-based models (GAT/TGNN) we use 8 attention heads. All GNNs use a hidden size of 768. We apply a normalization layer after each message-passing layer and do not use residual connections. Graph-level representations are obtained using one of \emph{attentional}, \emph{sum} or \emph{max pooling} , treated as a hyperparameter (Section~\ref{sec:training_setup}).


\textbf{PLM-only baselines :}
We evaluate two PLM-only baselines: \emph{frozen} and \emph{finetuned}.
In the frozen setting, we extract contextual token representations from the PLM's final hidden layer and compute a single vector by mean pooling over \emph{valid} tokens only (i.e., excluding padding and special tokens such as \texttt{[CLS]}, \texttt{[SEP]}, \texttt{<s>}, \texttt{</s>} depending on the tokenizer). Concretely, we apply attention-mask pooling and divide by the number of valid tokens. We then train a lightweight MLP classifier on top of this pooled representation.

For the finetuned setting, we unfreeze the PLM and jointly optimize its parameters together with the classifier head. To reduce memory usage during training, we employ gradient checkpointing and disable key–value caching when applicable. All PLMs are finetuned using mixed-precision training (bfloat16) with FlashAttention-2 enabled, and optimized with AdamW using a small learning rate ($1\mathrm{e}{-5}$). Class imbalance is handled via loss reweighting, and no additional regularization is applied beyond the classification head.

We consider three open-weight, code-specialized PLMs: DeepSeek-Coder-1.3B~\cite{guo2024deepseek},
StarCoder2-3B~\cite{lozhkov2024starcoder}, and Qwen2.5-Coder-0.5B~\cite{hui2024qwen2}.
This selection is motivated by two main factors. First, the models span a meaningful size range (0.5B, 1.3B, 3B), enabling a direct test of \textbf{RQ4}. Second, all models support relatively long contexts, which is compatible with our token budget. We include sliding-window handling only for completeness and for replication on longer inputs.

%

\textbf{PLM$\rightarrow$GNN hybrids :}
Hybrid models use frozen PLM token representations as semantic features that are aligned to AST nodes as explained in Section~\ref{sec:program_representation}. The resulting node semantic vectors are fused with the same structural features used by the GNN only baselines (node type features and positional features) and are fed to the GNN.
We freeze PLMs to avoid confounding architectural comparisons with large differences in finetuning stability, and to ensure all hybrid configurations can be trained under the same compute budget.

We instantiate three common GNN backbones:
GCN~\cite{kipf2016semi}, GAT~\cite{velivckovic2017graph}, and a neighborhood-restricted GraphTransformer (TGNN)~\cite{shi2020masked}. This selection provides a representative yet tractable suite of GNNs at varying levels of complexity. GCN is a canonical convolutional GNN that aggregates messages from neighbors providing a simple and widely used baseline. GAT refines this mechanism by learning neighbor-specific importance weights via multi-head attention, allowing the model to emphasize informative neighbors. Finally, TGNN introduces Transformer-style self-attention while restricting attention to local neighborhoods, offering a stronger attention-based baseline without incurring full-graph quadratic attention.

Pairing these backbones with different PLM feature sources directly supports \textbf{RQ4} by testing if the benefits of injected semantics depend on the GNN family used.

All GNN-only and PLM$\rightarrow$GNN hybrid models operate on the same AST graph structures and follow the same representation and training pipeline described in Sections~\ref{sec:program_representation} and~\ref{sec:training_setup}. The main difference is the inclusion of semantic features and the extractor used for such features.

\subsection{Program Representation and Feature Construction}
\label{sec:program_representation}

\paragraph{Graph construction.}
We parse each program/function into an Abstract Syntax Tree (AST) using the \texttt{Tree-sitter} parser via the \texttt{code-ast} library. We use the Java grammar for Java250 and the C grammar for Devign. All samples were parsed successfully; no instances were discarded due to parsing failures.

Edges encode parent--child and sibling relations. For message passing, we use bidirectional connectivity by adding reverse edges (equivalently, treating edges as undirected). We choose ASTs because they are widely supported and comparatively lightweight to construct. Previous work indicates that, for graph-based vulnerability detection, an AST-only pipeline retains most of the predictive signal~\cite{zhuang2021software}. Richer program graphs such as PDG/CPG can improve performance in some settings~\cite{zhou2019devign, paiva2024comparing}, but require heavier static analysis and increase preprocessing complexity and graph size, affecting reproducibility and cost.

The constructed AST graph is not fed directly into the PLM. Instead, the AST branch and the PLM branch both start from the same original source code string. We first parse the raw source code into an AST graph using \texttt{Tree-sitter}. In parallel, we feed the same raw source code text to the PLM tokenizer and encoder to obtain contextual token representations. We then align PLM token embeddings back to AST nodes using character-span overlap between tokenizer offsets and node spans. The resulting semantic vectors are attached to the corresponding AST nodes as node attributes, after which the GNN operates on the augmented graph.

\paragraph{Semantic feature extraction and node alignment.}
PLM feature extraction is performed from the raw source code text, not from the graph itself. For each program/function, we use the same source string that was parsed into an AST and pass it through a frozen PLM tokenizer and encoder to obtain contextual token representations. Let the tokenized input be $T=(t_1,\dots,t_L)$. If $L\le N_{\text{max\_tokens}}$, we process $T$ in a single forward pass. Otherwise, we use a sliding-window strategy: we split $T$ into $K$ overlapping fragments of length $l\le N_{\text{max\_tokens}}$ with stride $S$,
\[
F_i=\bigl(t_{(i-1)S+1},\dots,t_{(i-1)S+l}\bigr),\quad i=1,\dots,K
\]
For each fragment, the PLM outputs hidden states $H_i\in\mathbb{R}^{l\times h}$. We reconstruct $H\in\mathbb{R}^{L\times h}$ by merging window outputs and keeping a single representation per original token.

Each AST node $v$ corresponds to a contiguous character span $[c_v^{\text{start}},c_v^{\text{end}}]$ in the source. We identify tokens whose textual spans overlap this interval,
\[
I_v=\{j \mid t_j \text{ overlaps } [c_v^{\text{start}},c_v^{\text{end}}]\},
\]
and average their representations from the PLM's final layer to obtain a node semantic vector,
\[
h_v=\frac{1}{|I_v|}\sum_{j\in I_v} H_{j}\in\mathbb{R}^{h}.
\]
If an AST node aligns to no tokens (rare in practice), we set $h_v=\mathbf{0}$. We use mean pooling because it is deterministic and parameter-free; exploring learned pooling is left for future work. We obtain token offset mappings (start/end positions) from the tokenizer and node spans from \texttt{Tree-sitter}, converting spans to a common coordinate system before computing overlap.

All samples fit within context budgets in our experiments; we include sliding window for completeness and replication on longer inputs.

\paragraph{Structural node features.}
Following~\cite{dwivedi2023benchmarking}, we adopt Laplacian positional encoding. For each AST graph, we form the symmetric normalized Laplacian
\[
L_{\mathrm{sym}} = I - D^{-\tfrac12} A D^{-\tfrac12}
\]
and use the first $k$ non-trivial eigenvectors as a $k$-dimensional positional encoding $p_v$ per node. In this work, we set $k=32$.

Each AST node also has a discrete type ID $t_v$. We map $t_v$ to a learnable embedding $q_v\in\mathbb{R}^{d_{\text{type}}}$ via an embedding layer~\cite{bengio2003neural}.

\paragraph{Feature fusion.}
Each node $v$ has semantic features $h_v$, positional features $p_v$, and a node-type embedding $q_v$. We project each modality separately into a shared dimension $d_f$:
\[
\tilde{h}_v=\phi_{\text{sem}}(h_v),\quad \tilde{p}_v=\phi_{\text{pos}}(p_v),\quad \tilde{q}_v=\phi_{\text{type}}(q_v).
\]
We then fuse modalities using one of three strategies: concatenation, summation, or gated summation. For GNN-only baselines, we disable the semantic branch and fuse only positional and node-type features. We treat the fusion strategy as a hyperparameter, as described in Section~\ref{sec:training_setup}.

\subsection{Training, Tuning, and Experimental Setup}
\label{sec:training_setup}

\paragraph{Hyperparameter tuning.}
To ensure a fair comparison, we apply hyperparameter tuning to all model families evaluated in this study: GNN-only baselines, frozen PLM-only baselines, and PLM$\rightarrow$GNN hybrids. Because our experimental matrix spans multiple PLM$\times$GNN combinations, we use a budgeted tuning protocol that balances fairness and computational tractability.

For \textbf{GNN-only} and \textbf{PLM$\rightarrow$GNN hybrid} models, we tune hyperparameters on Java250 and reuse the selected configuration for Devign. This keeps the tuning budget comparable across the large grid of PLM$\times$GNN combinations and reduces task-specific overfitting of the tuning process. The selected hyperparameters are therefore chosen solely based on Java250 validation performance and then applied unchanged to Devign, except for an optional class-weighting flag used when training on Devign due to label imbalance.

For \textbf{frozen PLM-only} baselines, the only trainable component is a lightweight MLP classifier on top of pooled PLM embeddings. Since this tuning is computationally inexpensive, we tune this MLP separately on Java250 and Devign to avoid artificially weakening PLM-only baselines due to cross-task transfer.

Table~\ref{tab:hpo_space} summarizes the search space used for GNN-only and PLM$\rightarrow$GNN tuning. To keep capacity comparable across GNN backbones, we fix core model capacity and tune only lightweight architectural and optimization choices (normalization, activation, fusion strategy, pooling, dropout, weight decay, and learning-rate settings). Core GNN capacity is fixed to 8 message-passing layers and 8 heads for GAT and TGNN.

\begin{table}[t]
  \centering
  \caption{Hyperparameter search space for GNN-only and PLM$\rightarrow$GNN tuning. Core GNN capacity is fixed (8 layers and 8 heads for GAT/TGNN).}
  \label{tab:hpo_space}
  \renewcommand{\arraystretch}{1.45}
  \setlength{\tabcolsep}{2.5pt}
  \small
  \begin{tabularx}{\columnwidth}{llX}
    \toprule
    Category & Hyperparameter & Values or range \\
    \midrule
    \multirow{4}{*}{Architecture}
      & Normalization layer      & graph normalization; layer normalization \\
      & Nonlinearity             & ReLU; Leaky ReLU; GELU \\
      & Feature fusion strategy  & concatenation; summation; gated summation \\
      & Graph pooling            & attentional pooling; sum pooling; max pooling \\
    \addlinespace
    \multirow{6}{*}{Training}
      & Dropout rate             & uniform in $[0.00,\,0.10]$ \\
      & Weight decay             & log-uniform in $[10^{-6},\,10^{-3}]$ \\
      & Initial learning rate    & log-uniform in $[5\times10^{-6},\,10^{-4}]$ \\
      & Learning-rate schedule   & One-Cycle cosine schedule \\
      & Warm-up fraction         & uniform in $[0.05,\,0.15]$ \\
      & LR peak/initial ratio    & $\{100,\,50,\,10\}$ \\
    \bottomrule
  \end{tabularx}
\end{table}

For frozen PLM-only baselines, we tune the MLP classifier and optimizer hyperparameters using: hidden dimension $\in \{256,512,1024,2048\}$, depth $\in [1,5]$, dropout $\in [0,0.6]$, weight decay $\in [10^{-6},10^{-2}]$ (log-uniform), learning rate $\in [10^{-4},3\times10^{-3}]$ (log-uniform), and label smoothing $\in [0,0.1]$. We keep One-Cycle scheduler settings fixed (div\_factor=100, pct\_start=0.15, final\_div\_factor=$10^{4}$, cosine annealing) to limit the search dimensionality. We tune the hyperparameters of this MLP on both tasks: we optimize validation F1 for Java250 and validation AUPRC for Devign.

We use Optuna~\cite{optuna_2019} with a Tree-structured Parzen Estimator (TPE) sampler and Hyperband-style pruning. Hyperband is configured with reduction factor $\eta=3$, a maximum budget of 12 epochs per trial, and a minimum of 3 epochs before a trial becomes eligible for pruning. Trials periodically report the validation objective and configurations that fall sufficiently behind the current best are terminated early. We run up to 40 Optuna trials per (PLM, GNN) configuration.

\paragraph{Implementation and hardware.}
PLM finetuning baselines were run on Google Colab with NVIDIA A100 80GB GPUs. PLM feature extraction for the hybrid pipeline was run on Google Colab with NVIDIA A100 40GB GPUs. All GNN-only and PLM$\rightarrow$GNN hybrid training and evaluation were run on an AWS \texttt{g6.16xlarge} instance with 1$\times$ NVIDIA L4 Tensor Core GPU (24\,GiB VRAM).

We ran experiments with PyTorch 2.8.0 and PyTorch Geometric. We used PyTorch Lightning for training loops, TorchMetrics for evaluation, and Optuna for hyperparameter tuning. For PLM loading/tokenization we used the HuggingFace Transformers stack.

For PLM feature extraction and PLM baselines, we use the model's native maximum context lengths, i.e., $N_{\text{max\_tokens}}=32{,}768$ for Qwen2.5-Coder-0.5B and $N_{\text{max\_tokens}}=16{,}384$ for DeepSeek-Coder-1.3B-base and StarCoder2-3B. In our datasets, inputs fit within these budgets, so we did not use a sliding-window strategy. For GNN-only and PLM$\rightarrow$GNN hybrids, we train for 20 epochs on Devign with batch size 24 and for 40 epochs on Java250 with batch size 32. PLM finetuning uses batch size 2 for 8 epochs with gradient accumulation 8. Exact configurations and environment setup commands are released in the artifact repository.

\subsection{Evaluation Protocol}
\label{sec:evaluation_protocol}

\paragraph{Metrics and statistical analysis.}
We evaluate all models as supervised classifiers against the benchmark ground-truth labels described in Section~\ref{sec:datasets_and_tasks}. For Java250, each prediction is one of 250 program-class labels, and correctness is determined by whether the predicted class matches the ground-truth problem label. For Devign and Devign-OOD, each prediction is binary (\emph{vulnerable} or \emph{non-vulnerable}), and correctness is determined by whether the predicted label matches the benchmark vulnerability label.

We report standard classification metrics on the held-out test set. For all tasks, we report \emph{precision}, \emph{recall}, and \emph{F1}. For Devign and Devign-OOD, we additionally report the \emph{area under the precision--recall curve} (AUPRC), which is especially informative under class imbalance. Precision, recall, and F1 are computed as \emph{macro averages} over classes, so that each class contributes equally and performance is not dominated by majority classes.

We repeat each training run with three random seeds that affect parameter initialization, dropout, and mini-batch shuffling. We report results as mean $\pm$ standard deviation over the three seeds for each model and dataset. Finetuned PLM results are single-run due to cost and are reported as a reference point; we do not include them in statistical comparisons.

\paragraph{Decision-threshold calibration on Devign.}
Devign is imbalanced and the F1 score depends on the decision threshold, whereas AUPRC is threshold-free. For each model and seed, we calibrate a threshold $\tau^\star$ on the non-obfuscated validation set by maximizing F1, then apply it unchanged to both the standard and obfuscated test sets:
\begin{equation}
\tau^\star = \arg\max_{\tau \in \mathcal{T}} \mathrm{F1}\big(\mathbf{y}_{\text{val}}, \mathbb{I}[\hat{\mathbf{p}}_{\text{val}} \ge \tau]\big),
\end{equation}
where $\mathcal{T}$ is a threshold grid (or PR-curve thresholds). We report AUPRC as the primary metric and calibrated F1/precision/recall as secondary metrics.

\section{Results}
\label{sec:results}

\begin{table*}[ht]
  \centering
  \resizebox{\textwidth}{!}{%
    \setlength{\tabcolsep}{12pt}
    \small
    \begin{tabular}{c l c c c}
      \toprule
      Semantic Feature Extractor (PLM)
        & GNN Variant 
        & F1-score 
        & Precision
        & Recall \\
      \midrule
      \multirow{5}{*}{\shortstack{DeepSeek\\(\texttt{DeepSeek-Coder-1.3B})}}
                  & GAT              & 98.25\% (0.72\%)  & 98.28\% (0.71\%)  & 98.25\% (0.72\%) \\
                  & GCN              & 96.92\% (0.40\%)  & 96.96\% (0.40\%)  & 96.95\% (0.40\%) \\
                  & GraphTransformer & 97.97\% (0.95\%)  & 97.98\% (0.93\%)  & 98.00\% (0.95\%) \\
                  & Frozen           & 89.94\% (0.12\%)  & 90.59\% (0.12\%)  & 90.72\% (0.09\%) \\
                  & Finetuned        & 98.16\%  & 98.05\%  & 98.28\%  \\
      \addlinespace
      \multirow{5}{*}{\shortstack{StarCoder\\(\texttt{StarCoder2-3B})}}
                  & GAT              & 98.34\% (0.62\%)  & 98.37\% (0.61\%)  & 98.34\% (0.62\%) \\
                  & GCN              & 97.98\% (0.48\%)  & 98.02\% (0.47\%)  & 97.99\% (0.47\%) \\
                  & GraphTransformer & 97.90\% (0.83\%)  & 97.93\% (0.83\%)  & 97.92\% (0.81\%) \\
                  & Frozen           & 88.54\% (2.23\%)  & 89.32\% (2.04\%)  & 89.30\% (2.08\%) \\
                  & Finetuned        & 98.11\%  & 98.15\%  & 98.08\% \\
      \addlinespace
      \multirow{5}{*}{\shortstack{QwenCoder\\(\texttt{Qwen2.5-Coder-0.5B})}}
                  & GAT              & 97.90\% (0.08\%)  & 97.92\% (0.08\%)  & 97.91\% (0.07\%) \\
                  & GCN              & 97.55\% (0.26\%)  & 97.58\% (0.26\%)  & 97.57\% (0.26\%) \\
                  & GraphTransformer & 97.42\% (0.66\%)  & 97.45\% (0.65\%)  & 97.44\% (0.66\%) \\
                  & Frozen           & 82.89\% (2.67\%)  & 83.92\% (2.51\%)  & 83.69\% (2.21\%) \\
                  & Finetuned        & 98.28\%   & 98.45\%  & 98.12\% \\
      \addlinespace
      \multirow{3}{*}{\shortstack{None\\(no PLM features)}}
                  & GAT              & 85.59\% (0.22\%)  & 86.06\% (0.34\%)  & 86.06\% (0.34\%) \\
                  & GCN              & 83.28\% (0.42\%)  & 83.41\% (0.42\%)  & 83.44\% (0.40\%) \\
                  & GraphTransformer & 90.56\% (0.17\%)  & 90.63\% (0.17\%)  & 90.68\% (0.17\%) \\
      \bottomrule
    \end{tabular}
  }
  \caption{
  Java250 classification performance for PLM$\rightarrow$GNN hybrids and baselines
  (F1, precision, recall; mean ± standard deviation over 3 random seeds).}
  \label{tab:java250_results}
\end{table*}

\begin{table*}[ht]
  \centering
  \resizebox{\textwidth}{!}{%
    \setlength{\tabcolsep}{12pt}
    \small
    \begin{tabular}{c l c c c c}
      \toprule
      Semantic Feature Extractor (PLM) 
        & GNN Variant 
        & AUPRC     
        & F1-score 
        & Precision
        & Recall \\
      \midrule
      \multirow{5}{*}{\shortstack{DeepSeek\\(\texttt{DeepSeek-Coder-1.3B})}}
                  & GAT              & 66.98\% (0.55\%)  & 65.64\% (0.27\%)  & 51.85\% (0.19\%)  & 91.72\% (0.50\%) \\
                  & GCN              & 63.26\% (0.11\%)  & 64.48\% (0.18\%)  & 50.81\% (0.92\%)  & 90.70\% (3.53\%) \\
                  & GraphTransformer & 65.16\% (0.07\%)  & 65.14\% (0.21\%)  & 52.23\% (0.58\%)  & 88.86\% (0.79\%) \\
                  & Frozen           & 67.37\% (0.55\%)  & 66.84\% (0.37\%)  & 53.32\% (0.82\%)  & 90.33\% (0.98\%) \\
                  & Finetuned        & 68.12\%  & 65.61\%  & 53.48\%  & 87.57\% \\
      \addlinespace
      \multirow{5}{*}{\shortstack{StarCoder\\(\texttt{StarCoder2-3B})}}
                  & GAT              & 72.46\% (0.61\%)  & 65.91\% (0.58\%)  & 54.63\% (1.21\%)  & 85.44\% (4.58\%) \\
                  & GCN              & 71.07\% (0.46\%)  & 65.17\% (0.23\%)  & 54.27\% (0.84\%)  & 83.67\% (2.70\%) \\
                  & GraphTransformer & 72.28\% (0.12\%)  & 65.44\% (0.58\%)  & 54.75\% (0.39\%)  & 83.28\% (2.42\%) \\
                  & Frozen           & 69.80\% (0.56\%)  & 66.03\% (0.28\%)  & 53.59\% (1.24\%)  & 86.92\% (3.94\%) \\
                  & Finetuned        & 73.78\%  & 68.68\%  & 55.27\%  & 93.47\%  \\
      \addlinespace
      \multirow{5}{*}{\shortstack{QwenCoder\\(\texttt{Qwen2.5-Coder-0.5B})}}
                  & GAT              & 73.56\% (0.26\%)  & 65.59\% (0.27\%)  & 54.59\% (1.19\%)  & 84.14\% (3.40\%) \\
                  & GCN              & 73.02\% (1.14\%)  & 65.42\% (0.39\%)  & 53.05\% (1.17\%)  & 87.58\% (4.46\%) \\
                  & GraphTransformer & 73.80\% (0.56\%)  & 66.10\% (0.13\%)  & 55.07\% (0.76\%)  & 84.53\% (1.79\%) \\
                  & Frozen           & 69.97\% (0.38\%)  & 66.69\% (0.43\%)  & 53.48\% (0.50\%)  & 89.32\% (2.85\%) \\
                  & Finetuned        & 72.96\%  & 67.80\%  & 52.85\%  & 94.50\% \\
      \addlinespace
      \multirow{3}{*}{\shortstack{None\\(no PLM features)}}
                  & GAT              & 55.66\% (0.29\%)  & 62.56\% (0.35\%)  & 46.89\% (0.03\%)  & 97.02\% (1.61\%) \\
                  & GCN              & 51.80\% (0.44\%)  & 62.22\% (0.07\%)  & 46.07\% (0.11\%)  & 98.91\% (0.66\%) \\
                  & GraphTransformer & 51.16\% (0.15\%)  & 62.43\% (0.08\%)  & 46.13\% (0.12\%)  & 99.62\% (0.31\%) \\
      \bottomrule
    \end{tabular}
  }
  \caption{
  Devign vulnerability detection performance for PLM$\rightarrow$GNN hybrids and baselines (AUPRC, F1, precision, recall; mean ± standard deviation over 3 random seeds).
}
  \label{tab:devign_results}
\end{table*}

\subsection{RQ1: Effectiveness}
\label{sec:effectiveness}
On Java250 (Table~\ref{tab:java250_results}), all PLM$\rightarrow$GNN hybrids substantially outperform both GNN-only baselines and frozen PLM baselines. For example, the best hybrid configuration (GAT + StarCoder2-3B) reaches an F1 of 98.34\%, compared to 90.56\% for the best GNN-only model and 89.94\% for the best performing frozen PLM baseline. This pattern is consistent across all three PLMs; injecting PLM features into a GNN yields substantially higher performance than using either the PLM or the GNN in isolation.

On Devign (Table~\ref{tab:devign_results}), the picture is more nuanced. GNN-only models remain worse, but hybrids and frozen PLMs achieve similar macro-F1 after tuning the decision threshold on the validation set. Notably, AUPRC shows clearer differences: hybrids typically improve over their frozen PLM counterparts, except for DeepSeek where the frozen and GAT-based hybrid models overlap within standard deviation. Hybrids based on the smaller Qwen2.5-Coder-0.5B extractor achieve the highest AUPRC (up to 73.80\% $\pm$ 0.56\%), indicating a better precision--recall trade-off despite the smaller backbone. Finally, since hybrid hyperparameters are transferred from Java250 whereas frozen PLM MLPs are tuned per task, the Devign comparison is conservative for hybrids.

Overall, hybrids outperform GNN-only models and often improve AUPRC over frozen PLM baselines, while achieving comparable F1 at the validation-tuned operating point.


%
\subsection{RQ2: Efficiency}
\label{sec:efficiency}

To quantify the runtime cost of our hybrids, we separate (i) preprocessing (Table~\ref{tab:preprocessing_times}) from (ii) GNN inference (Table~\ref{tab:devign_inference_time}), and report wall-clock times on the Devign test split (2732 samples, batch size $=64$).

Table~\ref{tab:devign_inference_time} reports GNN inference times across all PLM$\rightarrow$GNN combinations. For a full pass over the test set, all hybrids run in roughly $10$--$12$ seconds, and differences between GNN architectures are small: for a fixed PLM, GAT, GCN, and GraphTransformer differ by well under two seconds. The choice of PLM extractor has a modest effect (Qwen2.5-Coder-0.5B is consistently the fastest, StarCoder2-3B the slowest), but overall the incremental cost of the GNN layers on top of cached PLM features is low. The times shown in Table~\ref{tab:devign_inference_time} assume cached PLM and positional embeddings; therefore, they report the total time spent on GNN inference in all PLM$\rightarrow$GNN combinations.

Table~\ref{tab:preprocessing_times} breaks down the preprocessing pipeline. AST construction is relatively cheap ($\approx 23$ seconds for the full test set), whereas Laplacian positional encodings dominate the structural overhead ($\approx 255$ seconds), exceeding even the cost of PLM feature extraction for DeepSeek (162 s) and StarCoder (220 s), and remaining comparable to Qwen2.5-Coder (107 s). In other words, in our current implementation, the main bottleneck in the hybrid pipeline is not the GNN itself but the spectral positional encoding step.

These measurements suggest two practical takeaways. First, the structural side of the pipeline can be made substantially cheaper by replacing Laplacian eigenvector embeddings with lighter positional schemes, without changing the overall hybrid architecture. Second, the choice of PLM affects both accuracy and cost: smaller models such as Qwen2.5-Coder-0.5B not only deliver strong AUPRC on Devign (Tables~\ref{tab:devign_results} and~\ref{tab:devign_obf_results}) but also reduce preprocessing and inference time compared to larger extractors. For this kind of hybrid design, compact PLMs therefore offer a more attractive performance–cost trade-off than larger backbones.


\subsection{RQ3: Robustness under Identifier Obfuscation}
\label{sec:robustness}
\begin{table*}[ht]
  \centering
  \resizebox{\textwidth}{!}{%
    \setlength{\tabcolsep}{12pt}
    \small
    \begin{tabular}{c l c c c c}
      \toprule
      Semantic Feature Extractor (PLM)
        & GNN Variant 
        & AUPRC     
        & F1-score 
        & Precision
        & Recall \\
      \midrule
      \multirow{5}{*}{\shortstack{DeepSeek\\(\texttt{DeepSeek-Coder-1.3B})}}
                  & GAT              & 62.83\% (0.53\%)  & 63.09\% (0.58\%)  & 51.82\% (0.24\%)  & 82.47\% (2.41\%) \\
                  & GCN              & 59.46\% (1.06\%)  & 61.06\% (1.78\%)  & 49.60\% (0.65\%)  & 82.12\% (8.26\%) \\
                  & GraphTransformer & 60.17\% (0.42\%)  & 63.06\% (0.82\%)  & 51.22\% (0.36\%)  & 84.25\% (2.13\%) \\
                  & Frozen           & 62.94\% (0.80\%)  & 64.74\% (0.43\%)  & 51.54\% (0.70\%)  & 87.82\% (2.93\%) \\
                  & Finetuned        & 63.73\%   & 65.20\%  & 50.06\%  & 93.47\% \\
      \addlinespace
      \multirow{5}{*}{\shortstack{StarCoder\\(\texttt{StarCoder2-3B})}}
                  & GAT              & 65.58\% (1.57\%)  & 63.58\% (1.03\%)  & 51.50\% (0.92\%)  & 85.57\% (5.56\%) \\
                  & GCN              & 63.51\% (0.46\%)  & 62.57\% (0.32\%)  & 50.78\% (1.00\%)  & 83.52\% (2.26\%) \\
                  & GraphTransformer & 64.86\% (1.03\%)  & 62.92\% (1.00\%)  & 52.02\% (1.07\%)  & 82.00\% (5.70\%) \\
                  & Frozen           & 63.45\% (0.69\%)  & 63.59\% (1.73\%)  & 51.45\% (1.26\%)  & 84.79\% (8.87\%) \\
                  & Finetuned        & 61.51\%  & 68.68\%  & 55.27\%  & 95.78\% \\
      \addlinespace
      \multirow{5}{*}{\shortstack{QwenCoder\\(\texttt{Qwen2.5-Coder-0.5B})}}
                  & GAT              & 69.90\% (0.65\%)  & 63.45\% (0.49\%)  & 52.47\% (0.74\%)  & 82.44\% (2.97\%) \\
                  & GCN              & 69.80\% (0.40\%)  & 63.68\% (0.45\%)  & 50.54\% (2.02\%)  & 89.18\% (7.34\%) \\
                  & GraphTransformer & 70.48\% (0.74\%)  & 63.66\% (0.29\%)  & 52.82\% (0.96\%)  & 82.46\% (3.41\%) \\
                  & Frozen           & 63.77\% (0.50\%)  & 62.11\% (1.63\%)  & 52.24\% (1.18\%)  & 77.68\% (7.24\%) \\
                  & Finetuned        & 64.61\%  & 65.61\%  & 49.47\% & 97.37\% \\
      \bottomrule
    \end{tabular}
  }
  \caption{
  Devign vulnerability detection under identifier obfuscation (OOD) for PLM$\rightarrow$GNN hybrids and baselines (AUPRC, F1, precision, recall; mean ± standard deviation over 3 random seeds).
}
  \label{tab:devign_obf_results}
\end{table*}

Under identifier obfuscation on Devign (Table~\ref{tab:devign_obf_results}), all model families suffer a performance drop, but the degradation in F1 is relatively uniform across hybrids and PLM-only baselines. This indicates that no family is dramatically more robust in terms of the operating point chosen by threshold tuning.

In contrast, AUPRC reveals clearer differences. Hybrids based on Qwen2.5-Coder-0.5B exhibit the smallest loss in AUPRC compared to their non-obfuscated counterparts, suggesting slightly better robustness in terms of precision--recall trade-off. Moreover, for both StarCoder and QwenCoder, the hybrid PLM$\rightarrow$GNN models lose less AUPRC than the corresponding frozen PLM baselines, indicating that injecting structural information helps preserve ranking quality under identifier perturbations.

These results suggest that robustness is primarily inherited from the semantic feature extractor: within a given PLM, the three GNN variants achieve very similar AUPRC values whose differences fall within one standard deviation, whereas changing the PLM has a larger effect. In other words, the choice of PLM matters more for robustness than the choice of GNN architecture, but GNN hybrids still provide a consistent advantage over using the PLM alone.

Finally, GNN-only models as shown in Table~\ref{tab:devign_results} remain the weakest across all settings, confirming that structural information by itself cannot compensate for the loss of semantic cues from identifiers and surrounding context.


\subsection{RQ4: Design Sensitivity}


\begin{table}[t]
\centering
\caption{Two-way ANOVA results for PLM$\rightarrow$GNN hybrid models on Java250; dependent variable: F1}
\label{tab:anova_java250}
\begin{tabular}{lrrr}
\toprule
Effect  &  $F$ & $p$-value & Partial $\eta^2$ \\
\midrule
GNN                    & 1.84352 & 0.186918 & 0.1701 \\
PLM                    & 0.91082 & 0.419949 & 0.0917 \\
GNN $\times$ PLM      & 0.69844 & 0.602963 & 0.1342 \\
\bottomrule
\end{tabular}
\end{table}


\begin{table}[t]
\centering
\caption{Two-way ANOVA results for PLM$\rightarrow$GNN hybrid models on Devign; dependent variable: AUPRC}
\label{tab:anova_devign}
\begin{tabular}{lrrr}
\toprule
Effect & $F$ & $p$-value & Partial $\eta^2$ \\
\midrule
GNN               & 19.223307 & 3.409664e-05 & 0.6811 \\
PLM               & 406.971014 & 1.038999e-15 & 0.9784 \\
GNN $\times$ PLM  & 4.850079  & 7.853120e-03 & 0.5187 \\
\bottomrule
\end{tabular}
\end{table}

\begin{table}[t]
\centering
\caption{Two-way ANOVA results for PLM$\rightarrow$GNN hybrid models on Devign under identifier obfuscation (OOD); dependent variable: AUPRC.}
\label{tab:anova_devign_obf}
\begin{tabular}{lrrr}
\toprule
Effect & $F$ & $p$-value & Partial $\eta^2$ \\
\midrule
GNN               & 7.320833  & 4.715218e-03 & 0.4485 \\
PLM               & 185.7346  & 9.620395e-13 & 0.9538 \\
GNN $\times$ PLM  & 2.641273  & 6.780997e-02 & 0.3698 \\
\bottomrule
\end{tabular}
\end{table}

Across settings, PLM identity is a major determinant of hybrid performance on Devign, while Java250 exhibits near-ceiling behavior where differences among hybrid design choices are small.

On Java250 (Table~\ref{tab:java250_results}), switching the PLM feature extractor changes F1 only marginally, and the gap between GNN backbones (GCN, GAT, GraphTransformer) stays within roughly one percentage point for a fixed PLM. Consistently, the two-way ANOVA (Table~\ref{tab:anova_java250}) detects no statistically significant main effects nor interaction, which we attribute to near-ceiling performance.

On Devign (Table~\ref{tab:devign_results}), the ranking differs: hybrids using Qwen2.5-Coder-0.5B achieve the strongest AUPRC despite relying on the smallest PLM. This suggests that larger PLMs are not necessarily better feature sources in our pipeline, and that performance depends more on representation quality than on parameter count alone. We note that this observation is based on three PLM backbones and does not isolate model size as an independent causal factor.

The two-way ANOVA on Devign hybrids (Table~\ref{tab:anova_devign}) confirms that both PLM and GNN choice significantly affect AUPRC, with PLM explaining the largest share of variance (partial $\eta^2\approx0.98$) and a significant in-distribution interaction (PLM$\times$GNN, $p=7.85\times10^{-3}$), indicating that the best GNN depends on the PLM feature source. Under identifier obfuscation (Table~\ref{tab:anova_devign_obf}), PLM remains dominant (partial $\eta^2\approx0.95$) and GNN remains significant ($p=4.72\times10^{-3}$), while the interaction weakens ($p=6.78\times10^{-2}$), suggesting reduced sensitivity to specific PLM$\rightarrow$GNN pairings under distribution shift.

Regarding the GNN backbone, GAT-based hybrids are most frequently best or tied across settings, though margins over GraphTransformer are typically small. Consequently, GAT is a reasonable default when extensive model selection is infeasible, while GraphTransformer can be competitive in PLM-specific cases. Finally, GNN-only models underperform substantially in both F1 (Java250) and AUPRC (Devign), confirming that structural signals alone are insufficient to match approaches that leverage pretrained semantic features.


\begin{table}[ht]
  \centering
  \caption{Inference time (in seconds) for GNN encoders on the Devign test set (2732 samples, batch size = 64). Each value is the wall-clock time for a full pass over the test split.}
  \label{tab:devign_inference_time}
  \small
  \begin{tabular}{lccc}
    \toprule
    \textbf{GNN Variant} & \textbf{DeepSeek} & \textbf{StarCoder} & \textbf{QwenCoder} \\
    \midrule
    GAT               & 11.30 & 12.06 & 10.13 \\
    GCN               & 10.85 & 11.41 & 10.65 \\
    GraphTransformer & 11.34 & 11.82 & 11.21 \\
    \bottomrule
    
  \end{tabular}
\end{table}

\begin{table}[ht]
  \centering
  \caption{Preprocessing time (in seconds) for each stage of the hybrid PLM$\rightarrow$GNN pipeline on the Devign test set. Times reflect a full pass over all 2732 samples.}
  \label{tab:preprocessing_times}
  \small
  \begin{tabular}{lc}
    \toprule
    \textbf{Preprocessing step} & \textbf{Time (s)} \\
    \midrule
    AST Construction                   & 23.33 \\
    Positional Feature Construction    & 254.72 \\
    DeepSeek Feature Extraction        & 161.98 \\
    StarCoder Feature Extraction       & 219.84 \\
    QwenCoder Feature Extraction       & 106.84 \\
    \bottomrule
  \end{tabular}
\end{table}

\section{Discussion}
\label{sec:discussion}
This section summarizes the main findings, practical guidance, and threats to validity.

\subsection{Key findings}
\label{sec:key_findings}

\textbf{RQ1 (Effectiveness) :}
On Java250, hybrids consistently outperform both GNN-only models and frozen PLM baselines by a large margin, indicating that AST structure and pretrained semantics are complementary in multiclass program classification. In contrast, on Devign, macro-F1 at the validation-tuned operating point is similar between hybrids and frozen PLMs, while AUPRC more clearly favors hybrids. A practical interpretation is that hybrids improve ranking quality (precision--recall trade-off) under imbalance, even when the optimal single-threshold F1 is similar.

\textbf{RQ2 (Efficiency) :}
The dominant cost in the hybrid pipeline is preprocessing, in particular the Laplacian positional encodings, which can exceed PLM feature extraction time on Devign. This highlights a non-obvious point: the expensive component is not necessarily the PLM forward pass, but can be the structural feature computation. Nevertheless, compared to end-to-end PLM finetuning, PLM$\rightarrow$GNN hybrids provide a substantially cheaper training alternative as they require less powerful hardware to train.

\textbf{RQ3 (Robustness) :}
Identifier obfuscation degrades all models, but hybrids based on Qwen2.5-Coder-0.5B exhibit the smallest AUPRC drop and generally outperform frozen PLM baselines under shift. Across hybrids, changing the PLM tends to matter more than changing the GNN backbone, suggesting that robustness is largely inherited from the semantic feature extractor. The observed AUPRC advantage of hybrids for StarCoder and Qwen is consistent with structure helping reduce reliance on identifier cues.

\textbf{RQ4 (Design sensitivity) :}
On Java250 differences are small, but on Devign the smallest PLM (Qwen2.5-Coder-0.5B) provides the best AUPRC as a feature source. Parameter count alone does not predict usefulness as an embedding source for downstream graph reasoning. A compact code-PLM can be both faster and more effective in the hybrid pipeline than larger alternatives. We hypothesize this can occur because frozen feature usefulness depends on representation properties rather than scale alone. Plausible factors include tokenization differences, embedding geometry, and pretraining-data/task alignment. We leave targeted diagnostics to future work.

\subsection{Practical guidance}
\label{sec:practical_guidance}
We translate our empirical results into actionable guidance for choosing between (i) frozen PLMs, (ii) PLM$\rightarrow$GNN hybrids, and (iii) finetuned PLMs, and for selecting a PLM feature source and GNN backbone under common constraints. These recommendations apply to the frozen-feature injection setting studied in this paper as we do not evaluate joint PLM+GNN finetuning.

\textbf{When to use a PLM$\rightarrow$GNN hybrid.}
Our results suggest that a hybrid is a strong option when you want much of the benefit of pretrained semantics while keeping training relatively practical and retaining structure-awareness.
On Java250, hybrids consistently outperform both GNN-only and frozen PLM baselines, and approach finetuned PLM performance (Table~\ref{tab:java250_results}).
On Devign, hybrids do not always yield the best calibrated macro-F1, but they more consistently improve ranking quality (AUPRC), which is often more informative under class imbalance (Table~\ref{tab:devign_results}).
This comes with added engineering and preprocessing overhead (parsing, alignment, positional encodings), so the gain is most justified when you can amortize preprocessing or cache PLM features (Section~\ref{sec:training_setup}).

\textbf{When a frozen PLM may be enough.}
If simplicity and minimal engineering dominate, frozen PLMs are a strong baseline: they avoid graph construction and alignment complexity and can be competitive in calibrated F1 on Devign.
However, our results show that frozen PLMs are clearly behind hybrids on Java250 and often behind hybrids in Devign AUPRC (Tables~\ref{tab:java250_results},~\ref{tab:devign_results}). This choice is therefore best seen as a simplicity–performance tradeoff.

\textbf{When to finetune the PLM.}
Finetuning is the right choice when peak performance is required and you can afford higher training cost and GPU memory requirements.
In our study, finetuning reaches near-ceiling performance on Java250 and is competitive on Devign, but it requires substantially more expensive hardware than training hybrids (Section~\ref{sec:training_setup}).

\textbf{Choosing the PLM feature extractor: do \emph{not} assume bigger is better.}
On Devign, the smallest extractor (Qwen2.5-Coder-0.5B) yields the best hybrid AUPRC and shows strong robustness under obfuscation (Tables~\ref{tab:devign_results},~\ref{tab:devign_obf_results}), despite being smaller than DeepSeek-1.3B and StarCoder2-3B.
The ANOVA confirms that PLM identity explains most variance in AUPRC on Devign (Tables~\ref{tab:anova_devign},~\ref{tab:anova_devign_obf}).
Practically: shortlist 2--3 PLMs, evaluate them first as \emph{frozen} baselines on the target task, and then prioritize the PLMs that already yield strong frozen performance as feature sources for hybrids. Parameter count alone is an unreliable proxy for usefulness as an embedding source in this pipeline.

\textbf{Choosing the GNN backbone: GAT is a strong default in our setting.}
Across datasets and settings, GAT-based hybrids are most frequently best or tied, while margins against the GraphTransformer are usually small (Tables~\ref{tab:java250_results},~\ref{tab:devign_results}).
GCN is consistently a weaker choice on Devign AUPRC and offers no compensating advantage in our results.
If you must pick one backbone without extensive selection, GAT is a reasonable default. If you can afford limited selection per PLM, the GraphTransformer can be competitive in PLM-specific cases as discussed in RQ4.

\textbf{Deployment cost is dominated by preprocessing, not GNN inference.}
With cached PLM features, the incremental inference cost of the GNN is small (Table~\ref{tab:devign_inference_time}).
In contrast, Laplacian positional encodings dominate preprocessing time in our implementation (Table~\ref{tab:preprocessing_times}).
If end-to-end latency or throughput matters, the highest-ROI optimization is to replace Laplacian eigenvectors with cheaper positional schemes (e.g., depth-to-root, degree features, random-walk features) or to disable positional encodings when acceptable.

\textbf{Robustness under identifier obfuscation: hybrids help, but the PLM matters most.}
All models degrade under identifier obfuscation, but hybrids based on Qwen2.5-Coder preserve AUPRC better than other extractors and outperform frozen PLM baselines under shift (Table~\ref{tab:devign_obf_results}).
Across hybrids, changing the PLM has a larger effect than changing the GNN, indicating that robustness is primarily inherited from the semantic extractor, with structure providing a consistent secondary benefit.

\textbf{A minimal decision rule :}
(i) If you can finetune and need the strongest in-distribution performance: finetune the PLM.
(ii) If you cannot finetune but want strong performance: use a PLM$\rightarrow$GNN hybrid with a compact, high-performing frozen extractor (in our study, Qwen2.5-Coder-0.5B) and GAT as the default backbone.
(iii) If simplicity dominates and you can accept weaker performance: use a frozen PLM baseline.

\subsection{Threats to validity}
\label{sec:threats_to_validity}

\subsubsection{Construct validity}
\label{sec:construct_validity}
Construct validity concerns whether our design and measurements capture the intended constructs: (i) predictive effectiveness, (ii) robustness to superficial cues, and (iii) computational efficiency.

\textbf{Effectiveness :} For Java250 we report macro-precision/recall/F1 to reflect balanced performance across 250 classes. For Devign (imbalanced), we treat AUPRC as the primary metric and report macro-F1 secondarily. Because F1 depends on a decision threshold, we calibrate the threshold on the validation set and apply it unchanged to the test set (and to Devign-OOD).

\textbf{Robustness : } We operationalize robustness via identifier obfuscation (Devign-OOD), which removes user-defined naming cues while preserving syntax and parseability. This tests reliance on identifier semantics, but represents only one axis of distribution shift.

\textbf{Efficiency : } We report wall-clock preprocessing time (AST parsing, positional encodings, PLM feature extraction) separately from GNN inference time assuming cached features; times are hardware/implementation dependent and are intended to support relative cost comparisons.

\subsubsection{Internal validity}
\label{sec:internal_validity}
Internal validity concerns whether the observed differences are caused by the modeled factors (PLM source, GNN backbone) rather than confounders in the training/evaluation pipeline. We mitigate this by using fixed dataset splits, a unified training budget and early-stopping protocol, identical graph construction and feature pipelines across methods, and three random seeds for each configuration. For Devign, we calibrate the decision threshold on the validation set and apply it unchanged to the test set and Devign-OOD, avoiding test-time tuning. Remaining threats include sensitivity to implementation and hardware details, and the fact that hybrid hyperparameters are transferred from Java250 to Devign, which may under- or over-favor some hybrids relative to per-task tuning.

\subsubsection{External validity}
\label{sec:external_validity}
External validity concerns generalization beyond the evaluated datasets, languages, and structural representations. Java250 is derived from competitive programming and may contain stylistic templates that differ from production code. Furthermore, Devign is limited to a specific collection of projects and provides function-level labels mined from commits, which may be noisy and does not capture inter-procedural vulnerabilities. Our graphs are AST-only (parent/child and sibling edges) and richer representations such as PDG/CPG may change accuracy and cost trade-offs. Finally, we evaluate three code PLMs and three GNN families; conclusions about model size or architecture families should be interpreted within this design space.

\section{Related work}
\label{sec:related}
We review prior work on code representation, organized into (i) sequence-based pretrained Transformers, (ii) structure-based graph models, and (iii) hybrid approaches that combine both.

Sequence-based pretrained language models (PLMs) have become the dominant paradigm for code understanding, motivated by the observation that software exhibits predictable statistical regularities that can be learned from large corpora~\cite{hindle2016naturalness, allamanis2018survey}. Following the success of Transformers~\cite{vaswani2017attention}, a large body of work has pretrained and finetuned Transformer models on source code for downstream tasks. Representative examples include encoder-style models (e.g., CodeBERT~\cite{feng2020codebert}) and encoder--decoder / text-to-text models (e.g., PLBART~\cite{ahmad2021unified}, CodeT5~\cite{wang2021codet5}), as well as more recent large code-specialized PLMs (e.g., CodeLlama~\cite{roziere2023code}).

Despite strong performance, these models process code primarily as a token sequence. As a result, structural information (e.g., AST relations) is not explicitly represented and must be learned implicitly from the sequence, leaving open when explicit structure can improve effectiveness or efficiency, especially under distribution shifts.

Graph-based representations make program structure explicit through artifacts such as abstract syntax trees (ASTs) and program graphs. Prior work has leveraged GNNs over program graphs for both generative and discriminative tasks. For example, Allamanis et al.\ proposed graph neural models over program graphs for code-related prediction tasks~\cite{allamanis2018learning}, and subsequent work applied AST- and graph-based models to summarization~\cite{leclair2020improved}, type inference~\cite{wei2020lambdanet}, code classification~\cite{zhang2022learning}, vulnerability detection~\cite{nguyen2022regvd}, and clone detection~\cite{zhang2024cross}. While structure-based models exploit syntactic relations directly, they typically lack the broad semantic priors obtained via large-scale pretraining. This motivates studying whether injecting pretrained semantic features into GNNs can close the gap to strong PLM baselines.


Hybrid approaches fuse token sequences with explicit code structure. GraphCodeBERT~\cite{guo2020graphcodebert} incorporates data-flow edges into a Transformer encoder via graph-guided masked attention. \cite{cheng2021gn} presents GN-Transformer which merges token and AST-node representations through a graph encoder and a Transformer decoder for summarization. AST-Trans~\cite{tang2022ast} linearizes ASTs for structure-aware attention, while SiT~\cite{wu2020code} uses adjacency masks to bias attention toward syntactically related tokens. CodeT5~\cite{wang2021codet5} and UniXcoder~\cite{guo2022unixcoder} inject structural signals (e.g., AST operations, path encodings, structural position embeddings) directly into pretraining.

Closer to our setting, recent work initializes GNN nodes with frozen code-PLM embeddings. GNN-Coder~\cite{ye2025gnn} uses a GNN over ASTs and control-flow graphs, with node features derived from a frozen Transformer encoder, to improve code search. Vul-LMGNN~\cite{liu2025vul} combines code LLM embeddings with graph-based representations for vulnerability detection over comprehensive program graphs.

These hybrid methods demonstrate that structure can complement pretrained semantics. However, they typically fix a single PLM$\rightarrow$GNN pairing and a limited set of tasks. We are not aware of a study that simultaneously  (i) sweeps across multiple GNN architectures and multiple modern PLMs, (ii) compares hybrids against both frozen and fully finetuned PLM baselines, (iii) evaluates robustness under identifier obfuscation, and (iv) quantifies performance–efficiency trade-offs. Our work fills this gap by providing an evidence-based comparison of GNN-only, PLM-only, and PLM$\rightarrow$GNN hybrid models for code understanding tasks.


\section{Conclusion}
We presented a systematic empirical study of PLM$\rightarrow$GNN hybrids for code classification and vulnerability detection. On code classification and vulnerability detection tasks, these hybrids consistently outperform GNN-only models and often improve ranking quality over frozen PLMs, while on Devign performance depends more on the PLM feature source than on the GNN backbone. We also find that compact PLMs can offer the best performance--cost trade-off, and that preprocessing---especially Laplacian positional encodings---dominates the one-off cost in our current implementation.

\section{Acknowledgments}

We thank Amazon Web Services (AWS) for supporting this project by
providing the computational resources used to run the experiments.

ChatGPT was used for editorial purposes to suggest alternative phrasing and improve the writing clarity by the authors. All technical content, interpretations, and final wording were reviewed and approved by the authors.

\bibliographystyle{ACM-Reference-Format}
\bibliography{bibliography}
\end{document}